\documentclass{article}

\usepackage{PRIMEarxiv}

\usepackage[utf8]{inputenc} 
\usepackage[T1]{fontenc}    
\usepackage{hyperref}       
\usepackage{url}            
\usepackage{booktabs}       
\usepackage{amsfonts}       
\usepackage{nicefrac}       
\usepackage{microtype}      
\usepackage{lipsum}
\usepackage{fancyhdr}       
\usepackage{graphicx}       
\graphicspath{{media/}}     

\title{Towards a Maturity Model for Systematic \\Literature Review Process
}

\author{
  Vinicius dos Santos \\
  University of São Paulo \\
  São Carlos, Brazil \\
  \texttt{vinicius.dos.santos@usp.br} \\
  \And
  Rick Kazman \\
  University of Hawaii \\
  Honolulu, USA \\
  \texttt{kazman@hawaii.edu} \\
  \And
  Rafael Capilla \\
  Rey Juan Carlos University \\
  Madrid, Spain \\
  \texttt{rafael.capilla@urjc.es} \\
  \And
  Elisa Yumi Nakagawa \\
  University of São Paulo \\
  São Carlos, Brazil \\
  \texttt{elisa@icmc.usp.br} \\
}

\begin{document}
\maketitle

\begin{abstract}
Systematic literature reviews (SLR) have been increasingly conducted in software engineering and they provide significant benefits in terms of summarizing the state of the research.  The process of conducting SLR is complex, involving several activities and consuming considerable effort and time from researchers. Researchers often skip or poorly conduct essential activities, which introduce threats to validity, resulting in lower-quality SLR. 
But researchers are often unaware of what they could do to mature their SLR process, thus improving the SLR quality.
The main goal of this paper is to introduce a maturity model for the SLR process named MM4SLR. 
To this end, we were inspired by well-known models like CMMI (Capability Maturity Model Integration). We first identified 39 key practices for SLR from the literature and grouped them into nine goals that were further grouped into five process areas. We then organized the process areas into five maturity levels which compose our model. Our proof of concept, applying the MM4SLR to four published SLR showed that the MM4SLR is suitable for appraising SLR and can identify important flaws in SLR quality.
MM4SLR can therefore support researchers in creating their SLR processes and selecting practices that could be adopted to mature their processes.
\end{abstract}


\section{Introduction}

The software engineering (SE) community has adopted Systematic Literature Reviews (SLRs)  as a technique to summarize evidence from primary studies and support researchers in outlining the state of the art of a given research topic~\cite{Kitchenham15}. An SLR provides important benefits, such as the possibility of dealing with information from different studies in an unbiased manner \cite{Niazi15}, production of auditable and repeatable results \cite{Budgen2018Reporting}, and the opportunity to identify research gaps, trends, and perspectives for future investigations~\cite{Kitchenham15}. A SLR involves different phases and activities that should be systematically followed to ensure the reliability and reproducibility of results.

The SE community has discussed the barriers to conduct and update SLR~\cite{Santos2021, dosSantos2024}, its main threats to validity~\cite{Ampatzoglou2019Identifying}, as well as other inherent problems like the enormous time and effort involved\cite{Felizardo2020Automating}. Some SLRs lack rigor, leading to several quality problems. Some of the problems are unintentional, for instance, the bias induced by the lack of expertise and experience in the domain or even in the SLR conduct~\cite{Riaz2010}. Other problems are related to the lack of resources (time, stakeholders, or tools) that lead researchers to ignore certain activities, which then reduce SLR quality. 
A poor quality SLR often cannot be replicated or updated, leading researchers to conduct a new SLR. To mitigate these problems several approaches have been proposed (including methods, techniques, and tools) to avoid bias and increase quality. For instance, tools have been proposed to automate some SLR tasks~\cite{Felizardo2020Automating}, such as selection of primary studies \cite{Watanabe2020}, support of SLR updates \cite{Mendes2020When}, or adoption of machine learning techniques (\cite{Felizardo2020Automating}). These approaches can contribute to improving SLR quality; nevertheless, they stand as point, not definitive, solutions to the lack of quality of SLR studies.

\textit{Sustainability} was introduced to the SLR context as a disruptive vision to deal with SLR problems from a more holistic perspective~\cite{Santos2021,dosSantos2024}. From the sustainability perspective, the lack of SLR quality is associated with three dimensions: (i) the social dimension is related to human aspects such as the experience of researchers with SLRs or poor knowledge sharing; (ii) the economic dimension refers to the lack of resources to conduct or update an SLR, mainly the time and effort of stakeholders; and (iii) the technical dimension refers to the immaturity of supporting tools that can lead researchers to poorly handle and manage results. Due to the problems found in these dimensions, researchers often left aside or poorly conducted essential activities, including those practices that could mitigate the threats to validity of the SLR. The process applied many SLRs is quite far from that proposed in the literature, and researchers often do not know which activities they should prioritize to mature their process and, consequently, improve their SLR quality.
This paper introduces a maturity model for the SLR process named MM4SLR. To achieve our goal, we first analyzed the literature to come up with 39 key practices (KPs) to mitigate threats to the validity of SLR studies. We then synthesized and grouped those KPs into nine specific goals (SGs). We also mapped these KPs into five process areas (PAs), which were organized into five maturity levels for the SLR process. Similarly to those found in models like CMMI\footnote{\url{https://cmmiinstitute.com/}} (Capability Maturity Model Integration)\cite{Chrissis2011}, those PAs aid researchers to comprehend the maturity of their SLR process and further improve it. We also conducted a proof of concept, applying MM4SLR to four SLRs found in the literature. Our results showed that this model is suitable for application in the SE context and capable of identifying important flaws in SLR quality.

The remainder of this paper is organized as follows: Section~\ref{sec:researchMethod} presents our research method. Section~\ref{sec:practicesGoalsProcessAreas} details the KPs, SGs, and PAs. Section~\ref{sec:maturityModel} presents MM4SLR and its five maturity levels. Section~\ref{sec:validation} presents the proof of concept. Finally, Section~\ref{sec:conclusion} outlines our conclusions and future work.

\section{Research Method}
\label{sec:researchMethod}

\begin{figure}[!h]
    \centering
    \includegraphics[width=0.55\linewidth]{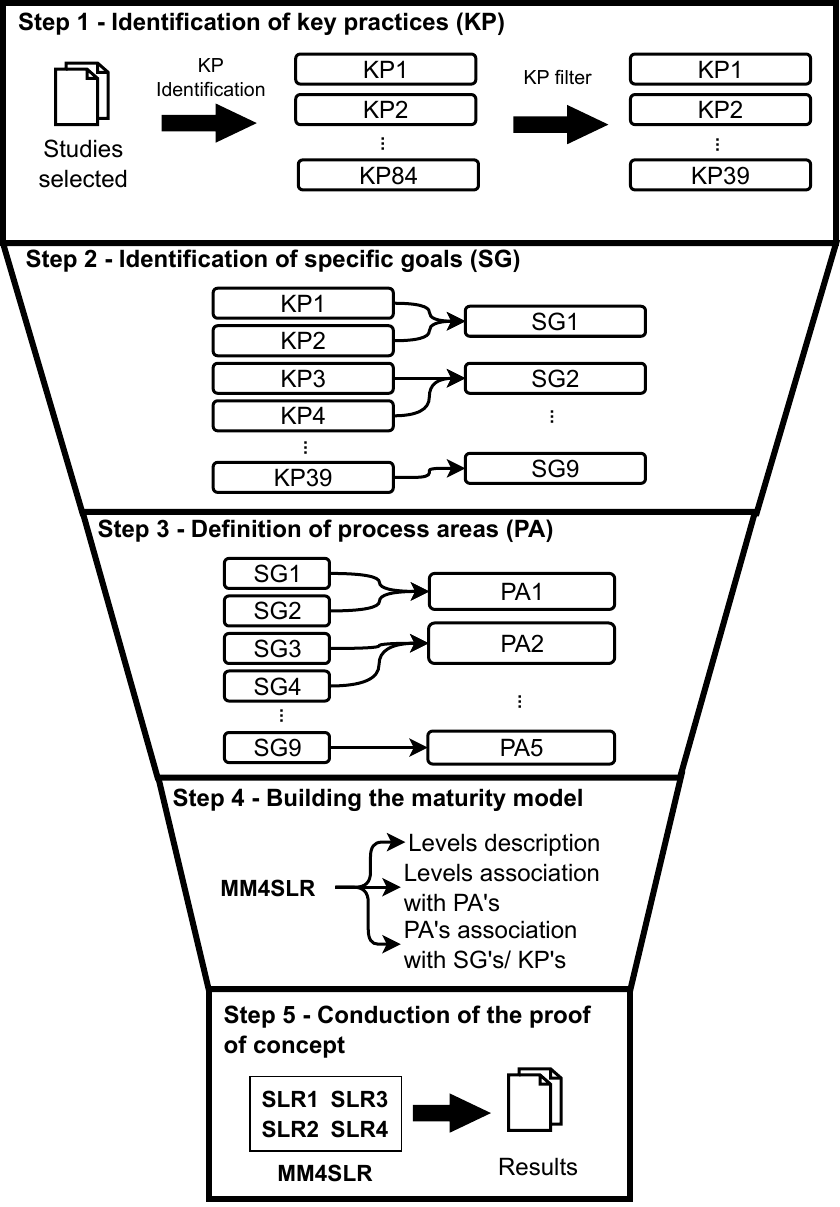}
    \caption{Research Method}
    \label{fig:method}
\end{figure}

Figure \ref{fig:method} presents the five-step process to build MM4SLR. Step 1 identified the KPs to mitigate threats to validity of SLRs using Thematic analysis~\cite{Cruzes2011b}. Step 2 grouped the KPs around similar SGs. Step 3 grouped the SGs into PAs that were inspired by the CMMI process \cite{Chrissis2011}. Step 4 used Grounded Theory (GT)~\cite{Anselm1998} to derive a new theory from the data and organize our maturity model. Finally, Step 5 conducted a proof-of-concept applying MM4SLR to four existing SLRs. Details about each step are presented as follows.

\noindent\textbf{Step 1. Identification of key practices (KPs):} We selected two well-cited and well-known tertiary studies \cite{Ampatzoglou2019Identifying,Zhou2016Threats} to extract KPs. Ampatzoglou et al. ~\cite{Ampatzoglou2019Identifying} analyzed 165 secondary studies published from 2007 to 2016, and Zhou et al.~\cite{Zhou2016Threats} assessed 316 studies published from 2004 to 2015 to identify the most common threats to the validity of the SLR process and strategies to mitigate them. We carefully examined the results of both studies and identified 84 candidate practices\footnote{The list of 84 practices is available on \url{https://anonymous.4open.science/r/MM4SLR-D43C/}} that could be adopted to mitigate the threats to validity in an SLR. We conducted group discussions and consensus meetings to synthesize 39 KPs that are mandatory, optional, or recommended, as discussed in Section~\ref{sec:practicesGoalsProcessAreas}. 

\noindent \textbf{Step 2. Identification of specific goals (SGs):} We grouped the 39 KPs into nine SGs (detailed in Section \ref{sec:practicesGoalsProcessAreas}). To do this, we clustered similar practices that addressed the same threat to validity. For instance, several practices can be adopted to ensure good coverage of an SLR (i.e.,  all relevant primary studies were considered or snowballing techniques were employed). The nine goals defined are critical points in the SLR process that may deserve more attention. 

\noindent\textbf{Step 3. Definition of process areas (PAs):} We clustered the SGs that addressed common issues, resulting in five PAs (detailed in Section~\ref{sec:practicesGoalsProcessAreas}). For example, those SGs associated with the definition of strategies for searching studies were assigned to the PA of Project Planning (PP). Hence, these PAs represent broader topics that, once satisfied, could mature the SLR process. 

\noindent\textbf{Step 4. Building the maturity model:} We analyzed the PAs and, inspired by the CMMI,  built MM4SLR that encompasses five levels of maturity for the SLR process.

\noindent\textbf{Step 5. Conduct of the proof of concept:} We applied MM4SLR to four existing SLRs to assess the degree to which our model could evaluate the maturity of each SLR process (detailed in Section \ref{sec:validation})

\section{Key Practices, Specific Goals, and Process Areas}
\label{sec:practicesGoalsProcessAreas}

\begin{table*}
    \centering
    \caption{Key practices for conducting an SLR}
    \includegraphics[width=1\linewidth]{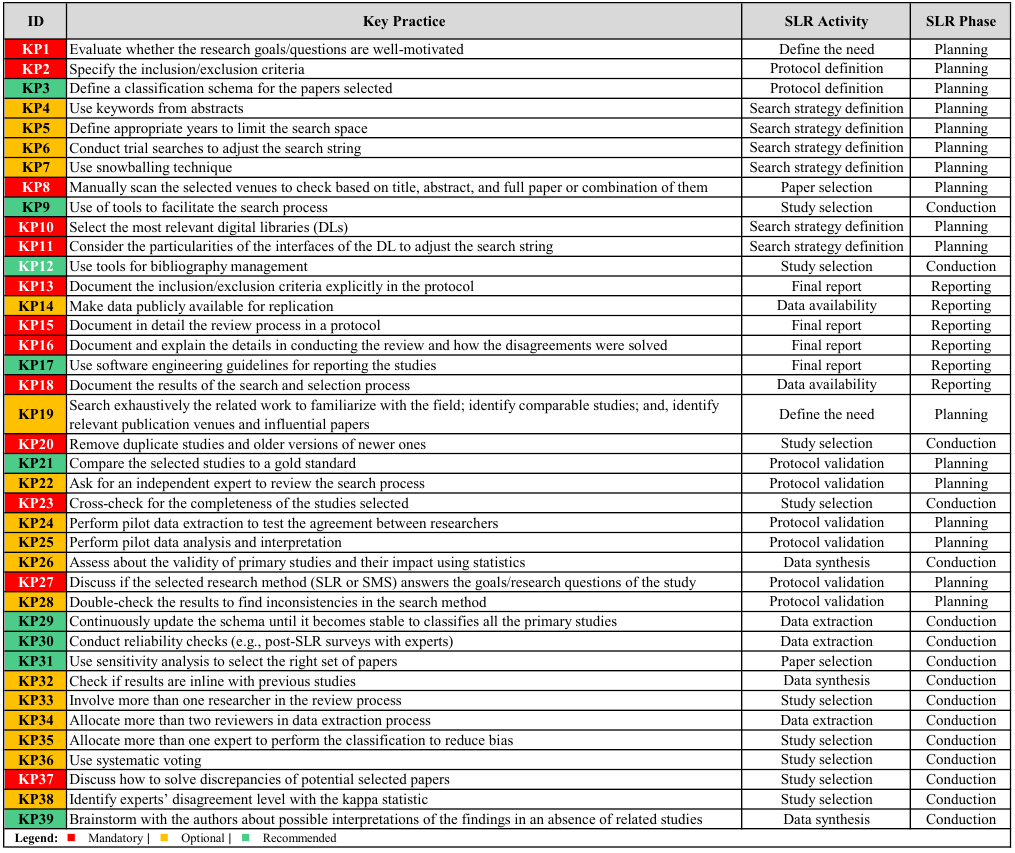}
    \label{tab:kps}
\end{table*}

Table \ref{tab:kps} presents the 39 KPs;  
rows in red represent mandatory activities when conducting an SLR,  green represents optional activities, and yellow represents recommended practices. We then mapped the KPs to nine SGs, as shown in Table \ref{tab:specificGoals}. We also associated the SGs with five PAs (i.e., Project Planning (PP), Technical Support (TS), Results Documentation (RD), Process and Product Quality Assurance (PPQA), and Communication Management (CM)), which appear in the last column of Table \ref{tab:kps} and are discussed below.

\begin{table}
\caption{Specific goals identified from key practices and associated process areas}
\label{tab:specificGoals}
\centering
\includegraphics[width=0.60\linewidth]{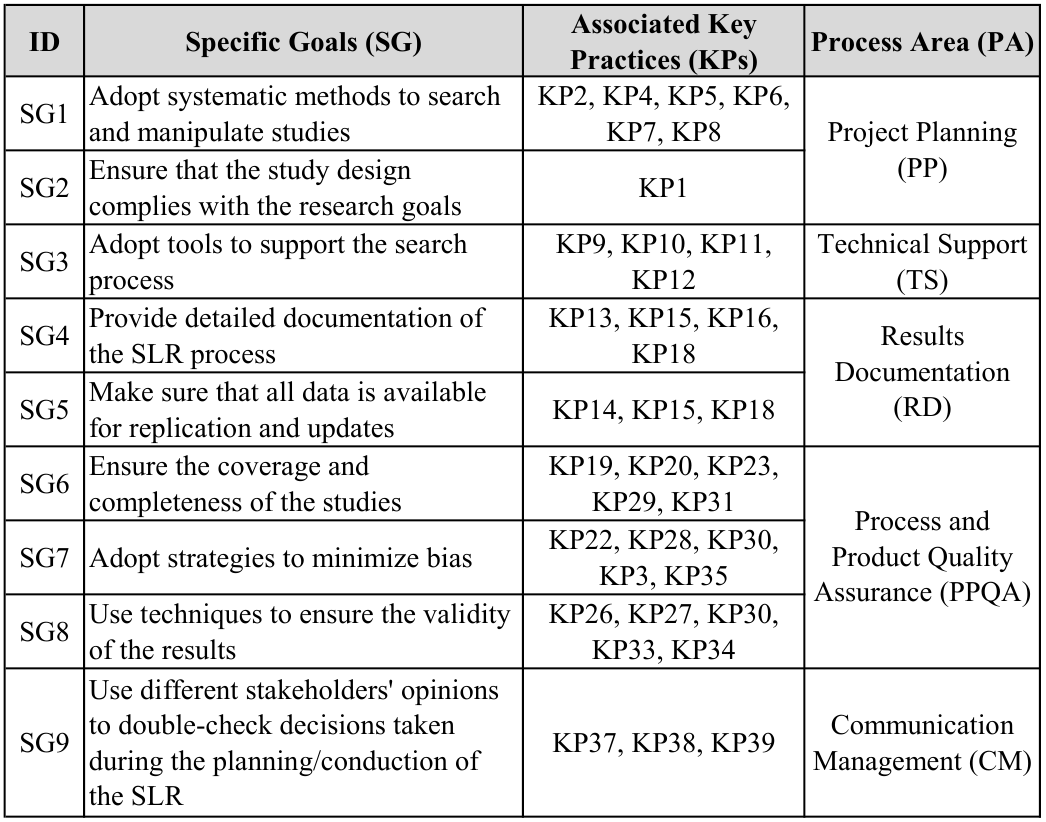}
\end{table}

\textit{Project Planning} (PP)  covers the entire SLR planning process, which defines how the review will be conducted. SLR planning concerns include the need for a review, the amount of resources needed (including budget, time, and human resources), and the review protocol. Planning an SLR is critical because decisions at this stage significantly impact the entire process. We identified eight practices (see Table \ref{tab:kps}) associated with these goals. We marked three out of eight as mandatory activities or Key Processes (KP1, KP2, and KP8) that any SLR must fulfill, one as optional (KP3), and the other four as recommended practices. Please note that tools could help to analyze the selected venues in practice area KP8 but in most cases, manual intervention is needed. Some of the recommended practices are done in many SLRs, which can improve the quality of the studies. However, in other cases where low-quality SLRs are produced, some of these recommended practices are done incorrectly (e.g., bad selection of studies) or simply neglected (e.g., conduct of a pilot or trial search to adjust the search string). Finally, KP3 was marked as  optional; it could be categorized as recommended but we found that some SLRs do not provide such classification. The snowballing technique (KP7) ensures the completeness of the studies selected but, in many cases, this task is not performed or the number of snowballing iterations is not enough. Overall, this PA can guide researchers in planning their reviews, avoiding common pitfalls, and offering directions to achieve results that impact their target audience positively.

\textit{Technical Support} (TS) is another PA that refers to the adoption of supporting tools and search engines (of databases) during the conduct and update of SLR. Hence, TS is essential to mitigate several problems such as high-time consumption. Carver et al. \cite{Carver2013} highlighted that the most time-consuming task is the search for studies in the databases, selection of studies, data extraction, and assessment of the quality of studies. Several tools have been proposed to support some of these tasks 
but most researchers use spreadsheets and reference managers to organize the results and remove duplicate entries \cite{AlZubidy2018}. In this light, specific tools supporting the SLR activities are still immature
and more efforts are needed to enhance and integrate them~\cite{Felizardo2020Automating}.  We identified four KPs as shown in Table \ref{tab:kps}. We categorized two of them as mandatory (KP10 and KP11),  while the other two (KP9 and KP12) are optional. We did not include in the discussion tools providing advanced visualization capabilities as most of them are still immature research tools.

\textit{Results Documentation} (RD) is the PA that provides ways to create reliable and complete documentation of the SLR, so that SLR results effectively impact the target audience. To create such documentation, we identified six practices: documenting in detail the whole process, providing access to the raw data (for anyone who intends to replicate or update the SLR), and documenting the details of the SLR process. However, the SE community has noticed the lack of impact of the SLR results on SE practitioners so they proposed new strategies for SLR documentation such as evidence briefings~\cite{Cartaxo2018Role} aiming to impact industry. From the six KPs shown in Table \ref{tab:kps}, four of them were classified as mandatory as SLR researchers need to publish all the details of the study for replication purposes and the techniques used. Only one key practice (KP14) was considered as recommended because not all SLRs make available the data used and another KP (KP17) was marked as optional as we did not find an SLR using SE guidelines.

\textit{Process and Product Quality Assurance} (PPQA)  encompass 17 practices (see Table \ref{tab:kps}) focused on improving the quality of the SLR. These practices aim to mitigate bias in the SLR process, such as handling inconsistencies, ensuring the validity of results, ensuring the reliability of the venues and studies analyzed, and avoiding bias in data extraction among others. Many of these KPs require significant additional effort which is why most of them are marked as recommended. For instance, we can perform pilot studies to adjust the search process using different keywords. Moreover, involving several researchers to resolve disagreements or cross-check results is sometimes not easy to accomplish. In addition, we identified three mandatory KPs (KP20, KP23, and KP27) as tasks that must be always performed to achieve a minimum level of quality, while the other four  (KP21, KP29, KP30, and KP31) were considered optional. For instance, we could use sensitivity analysis to know if the set of papers selected is the right ones. As more recommended practices are adopted the SLR will have better quality and less bias.

\textit{Communication Management} (CM) is the last PA,  aimed at handling the communication among the researchers involved in the SLR.  Communication is required to avoid misunderstandings in the previous KPs. From our analysis, we distilled four KP that may improve communication and solve conflicts among researchers. For instance: using systematic voting, adopting kappa statistics, brainstorming possible interpretations of findings, and cross-checking data extracted from studies. CM provides the means to gather and report insights from different perspectives, making the contribution of the SLR stronger.
As shown in Table \ref{tab:kps}, we identified one mandatory practice (KP37), two recommended ones (KP36 and KP38) that can improve the quality of the communication process, and one optional (KP39) that rarely happens.

\textbf{Specific Goals and Process Areas:} From the classification and clustering of the aforementioned 39 KPs, we derived a set of SGs,  as shown in Table \ref{tab:kps}. For instance, the Project Planning (PP) PA addresses goals SG1 and SG2 aimed to adopt systematic methods and ensure the study complies with the needs and research goals. The PA Technical Support (TS) involves goal SG3 and the use of tools to facilitate the conduct of the SLR. For SG3, we know two KPs are optional but any tool used to reduce the human burden to analyze the data is welcome.  SG4 and SG5 share some of the same KPs. With these two goals, we wanted to highlight that all the data and the steps of the protocol used must be publicly available. However, for replication or update purposes, the practice area KP16 was not included in SG5 as it can be different if the SLR researchers are different authors. 

The PA Process and Product Quality Assurances (PPQA) have three goals. 
In SG6, we address the coverage and completeness of the studies selected and represented by the KPs shown in Table \ref{tab:kps}. Snowballing can help in this goal to identify missing studies. Additionally, we felt it important to adopt strategies to minimize the bias (SG7) of the studies reviewed and ensure the validity of the results (SG8). Both goals are crucial because, for SG7, sometimes we discuss studies that are not relevant for the research questions addressed or we select studies that do not fit well under the inclusion/exclusion criteria. Furthermore, goal SG8 must be supported by valid studies selected (KP26) to derive meaningful results using the right studies (KP27), but reliability checks (KP30) can be helpful in cross-checking the initial results. Such cross-checks are often performed by more than one expert on the topic (KP33 and KP34). Our last goal (SG9) addresses communication management issues that are supported by the three KPs shown in Table \ref{tab:kps}, which bring experts to discuss possible discrepancies and interpretations of the results. We excluded KP36 (i.e., systematic voting) as there are several forms to achieve a consensus.

\section{A Maturity Model for the SLR Process}
\label{sec:maturityModel}

This section presents the initial version of MM4SLR, which resulted from the analysis of the KPs, SGs, and PAs and which was inspired by the CMMI maturity levels. We relied on CMMI 2.0 which defines five maturity levels (numbered from 1 to 5): incomplete, performed, managed, defined, quantitatively managed, and optimizing. 

\vspace{.2cm}\noindent \textbf{Level 0 (a.k.a. Incomplete)}: We did not include any PA for this level as it lacks any systematization and will produce SLRs with no rigor in the process or results. 

\vspace{.2cm}\noindent\textbf{Level 1 - Performed}: This level is concerned with the management of the SLR process. This level addresses two PAs: PP (which deals with SLR planning), and RD (which focuses on SLR documentation) making the review process managed and controlled with standardized procedures. At this level, the quality of the SLR is unknown as it depends on the specific KPs enacted and the reliability of the results is questionable.

\vspace{.2cm}\noindent\textbf{Level 2 - Managed}: SLR process is well-defined and documented and the focus of this level is to assure SLR quality. The main difference is the Technical Support (TS) that is used to manage the planning, conduct, and reporting of the SLR. At this level, there is a deeper comprehension of quality control mechanisms, but the SLR quality at this level is still a challenge.

\vspace{.2cm}\noindent\textbf{Level 3 - Defined}:  SLRs are conducted proactively making every effort to continuously improve their quality. To fulfill the requirements of this level, the SLR process must address two PAs: PPQA (which refers to quality assurance) and CM (which deals with communication). The SLR process assures the SLR quality, and biases are minimized in handling disagreements. Many practices refer to double-checking SLR activities or running assessments, such as the quality appraisal of primary studies and sensitivity analysis. CM is also assigned to this level because it is a well-defined activity that enhances the communication between SLR researchers and other stakeholders.

\vspace{.2cm}\noindent\textbf{Level 4 - Quantitatively Managed}: SLR process is quantitatively managed by metrics that ensure a systematic control of quality and effort applied in each activity. This level prioritizes the use of metrics, for instance, the sensitivity formula proposed by Zhang et al.~\cite{Zhang2010} that is defined as the proportion of studies retrieved by the search strategy adopted and the number of relevant studies retrieved. Specifically, statistical analysis of the results becomes essential and the impact of each primary study can be measured. 

\textbf{Level 5 - Optimizing)} of an SLR process is hard to achieve and refers to the highest level of maturity in which researchers are concerned with continuously evaluating and improving their SLR process. At this level researchers already overcome most of the SLR process challenges and look for opportunities to optimize the process to consume fewer resources and increase quality, for instance, adopting better SLR supporting tools. 
%

\section{Proof of concept}
\label{sec:validation}

We selected four SLRs to determine the viability of our model; in particular we analyzed how the SLRs were conducted (as reported in the papers), aiming to assess the processes used to conduct them.
We selected SLR1 and SLR2 from two top venues ranked in the first quartile (Q1) while SLR3 and SLR4 were published in other journals non-ranked in the Journal Citation Reports. We anonymized them to protect the identities of the authors. 

We applied MM4SLR following the two criteria: (i) the SLR should have all mandatory KPs of each PA; and (ii) the SLR should have at least 50\% of the KPs of each PA. If the SLR fulfilled such criteria, it is deemed to have fulfilled that PA. For instance, SLR1 has all the mandatory KPs of PA Project Planning (PP) and has 5 out of 8 KPs of this PA. SLR1 then fulfills the PA Project Planning (PP). After the evaluation, we can observe in Table \ref{tab:kps} that this SLR fulfills the following PAs: PP, TS, and RD. In the case of PPQA, there is one mandatory KP (KP23) for which we could not find clear evidence so we excluded it, and likewise with the PA Communication Management (CM). Consequently, according to MM4SLR, we can state that SLR1 meets Level 2 (Managed) but it requires more KPs to achieve Level 3. Regarding SLR2, we mapped to the same maturity level with some small variations in the recommended and optional KPs compared to SLR1. 

The case of SLR3 is a bit different. We rated it with Level 0 (Incomplete) because, from our evaluation, there is one mandatory KP (KP16) not fulfilled so this SLR cannot be in Level 1 (Performed). As the authors did not discuss how the disagreements in reviewing and classifying studies was done, we believe that the SLR process did not reach that maturity level. Although one might argue that it is only one mandatory KP not addressed, this KP is important to avoid bias in the selection and classification of the final set of studies. Finally, we rated SLR4 as Level 2 (Managed) as it fulfills the mandatory KPs for that maturity level. This SLR has other interesting KPs belonging to the PA PPQA that other studies do not have (e.g., KP36, voting system).

\begin{table}
    \centering
    \caption{Analysis of SLR using MM4SLR}
    \includegraphics[width=0.50\linewidth]{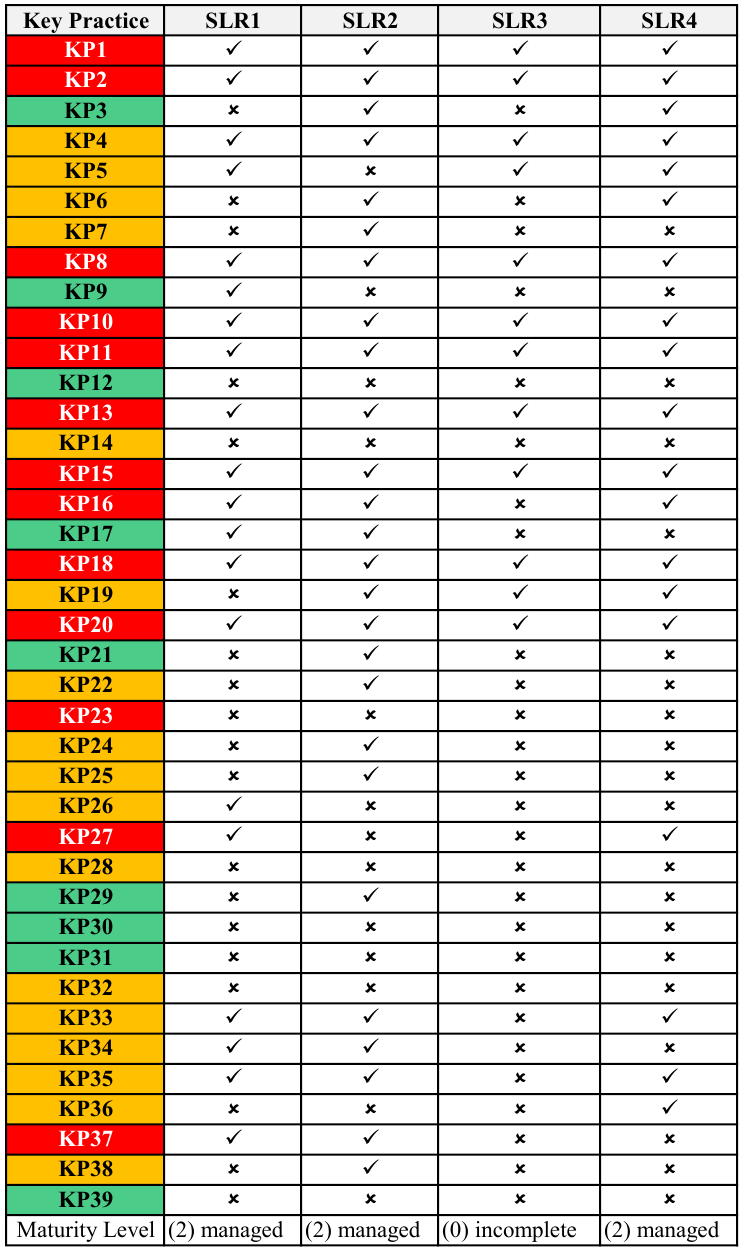}
    \label{tab:poc}
\end{table}

An increasing number of SLRs are being conducted  without concern for the quality of their process and product (i.e., SLR results). Despite the existence of well-known guidelines and processes to conduct SLRs~\cite{Kitchenham15}, many researchers do not follow them systematically. From our experience, it is not difficult to find SLRs published that are at maturity levels 1 or 2, which may produce results of lower quality. Probably, most of SLR that are evaluated by specialists belong to Levels 2 or 3, meaning that their authors are concerned with the quality of the SLR and results as well. SLR processes and KP in Level 4 are more uncommon but reachable. Our MM4SLR helps researchers evaluate and understand which KPs are  missing if they want to achieve a higher maturity level and thus what effort will be required. Knowing this, researchers can roughly estimate how much time this could take and how much effort is needed. This may also depend on the tool support used, the number of studies to analyze, or if snowballing is used. 

We do not want to introduce extra complexity or additional tasks in typical SLR processes but rather make the SLR process more understandable and highlight the quality aspects that make the difference between poor and good SLRs. In addition, we expect to verify the adequacy and distribution of the proposed SGs and KPs in each PA as some KPs  belong to more than one PA. One crucial point for discussion is to agree on how many KPs must be considered for a given SLR to achieve a particular maturity level. In our proof of concept, we relaxed the number of KPs to 50\% but each organization could decide on what this percentage should be and tune the proposed rating, specifically in those PAs that encompass more recommended items such as PPQA.

\section{Conclusion and Future Works}
\label{sec:conclusion}
This paper introduced a maturity model---MM4SLR---to help assess and improve current practices for SLR conduct. MM4SLR is the first attempt to guide researchers in improving their SLR processes. MM4SLR suggests mandatory, recommended, and optional activities (KPs) in the SLR process. We encourage addressing as many KPs as possible to ensure the SLR's completeness, accuracy, and validity. We recognize that fulfilling all KPs is difficult, so we recommend prioritizing those that could contribute to the quality of SLR.  

We suggest the following topics for future work: (i) additional validation of MM4SLR considering other SLRs; (ii) application of MM4SLR to increase the maturity level of SLR processes and the quality of resulting SLRs (iii) advice for how to more easily achieve Levels 4 and 5; (iv) rules to define the number of KPs and which ones are required to achieve a given maturity level; and (v) refinement of the MM4SLR with new KPs including the update or removal of existing ones based on new goals. 

Finally, many SLR activities are still conducted manually so we believe introducing machine learning-based solutions could considerably support this process, for instance, the reading of studies during studies selection, the conduct of snowballing, or data summarization during the synthesis of results. But this, too, comes with threats to validity which must be addressed.

\section*{Acknowledgments}
This work was supported by (2019/23663-1, 2023/00488-5, 2024/00329-7) and CNPq (313245/2021-5).

\bibliographystyle{unsrt}  
\bibliography{references}

\end{document}